\documentstyle[twoside,fleqn,espcrc2,epsfig]{article}

\def\be{\begin{equation}}
\def\ee{\end{equation}}
\def\bea{\begin{eqnarray}}
\def\eea{\end{eqnarray}}
\newcommand{\bb}{\bibitem}

\newcommand{\AmS}{{\protect\the\textfont2
  A\kern-.1667em\lower.5ex\hbox{M}\kern-.125emS}}

\title{Lattice calculation of the strangeness and electromagnetic nucleon
form factors}

\author{
            A.G.\ Williams
                 \address{CSSM and Department of Physics and Mathematical
                             Physics,
                      University of Adelaide, Australia 5005} 
                 \thanks{{\em e-mail:}~awilliam@physics.adelaide.edu.au,
                         \hfill\break
                         {\em URL:}~http://www.physics.adelaide.edu.au/cssm}
        }

\begin{document}

\begin{abstract}
We report on recent lattice QCD calculations
of the strangeness 
magnetic moment of the nucleon and the
nucleon electromagnetic form factors, when we allow the electromagnetic
current to connect to quark loops as well as to the valence quarks. Our
result for the strangeness magnetic moment is
$G_M^s(0) =  - 0.36 \pm 0.20 $. The sea contributions from the u and d
quarks are about 80\% larger. However, they cancel to a large extent
due to their electric charges, resulting in a smaller net sea contribution
of $ - 0.097 \pm 0.037 \mu_N$ to the nucleon magnetic moment. As far as the
neutron to proton magnetic moment ratio is concerned, this sea contribution
tends to cancel out the cloud-quark effect from the Z-graphs and result
in a ratio of $ -0.68 \pm 0.04$ which is close to the SU(6) relation and
the experiment.
The strangeness Sachs electric mean-square radius $\langle r_s^2\rangle_E$
is found to be small and negative and the total sea contributes
substantially to the neutron electric form factor.

\end{abstract}

\maketitle

We summarize some recent results \cite{dlw} on nucleon electromagentic form
factors, including the strangeness electric and magnetic form factors.
The strangeness content of the nucleon has been a topic of considerable
recent interest for a variety of reasons. The studies of nucleon
spin structure functions in polarized deep inelastic
scattering experiments at CERN and SLAC \cite{DIS}, combined with
neutron and hyperon $\beta$ decays, have turned up a surprisingly
large and negative polarization from the strange quark. In addition,
there is a well-known long-standing discrepancy between the
pion-nucleon sigma term
extracted from the low energy pion-nucleon scattering~\cite{gls91}
and that from the octect baryon masses~\cite{cheng76}. This
discrepancy can be reconciled if a significant $\bar{s}s$ content
in the nucleon~\cite{cheng76,gl82} is admitted.

To address some of these issues, an experiment to
measure the neutral weak magnetic form factor $G^Z_M$
via elastic parity-violating electron scattering 
was recently carried out
by the SAMPLE collaboration~\cite{SAMPLE97}. The strangeness magnetic form
factor is obtained by subtracting out the nucleon magnetic form
factors $G^p_M$ and $G^n_M$. The reported value is
$G^s_M(Q^2=0.1$GeV$^2)= +0.23\pm 0.37\pm 0.15\pm 0.19 $.
Future experiments have the promise of tightening the errors and isolating
the radiative corrections so that we can hope to have a well-determined
value and sign for $G^s_M(0)$.

Theoretical predictions of $G^s_M(0)$ vary widely. The values
from various models and analyses range from $ -0.75\pm 0.30 $
in a QCD equalities analysis~\cite{lei96} to $+ 0.37 $ in an $SU(3)$
chiral bag model~\cite{hpm97}. While a few give positive
values~\cite{hpm97,gi97}, most model predictions are negative with a
typical range of $-0.25$ to 
$-0.45$.
Summaries of these predictions can be found in
Refs.\ \cite{lei96,cbk96}. 
A similar situation exists for the
strangeness electric mean-square radius $\langle r_s^2\rangle_E$.
A number of the predictions are positive, while the
others are negative.
Elastic $\vec{e}\; p$ and $\vec{e}\;{}^{4}He$ parity-violation experiments
are currently planned at TJNAF~\cite{pve91} to
measure the asymmetry $A_{LR} $ at forward angles to extract
$\langle r_s^2\rangle_E$. Hopefully, they will settle
the issue of its sign.
 
In view of the large spread of theoretical predictions for both
$G^s_M(0)$ and $\langle r_s^2\rangle_E$ and in view of the fact
that the experimental errors on $G^s_M(0)$ are still large, it is clearly
important to perform a first-principles lattice QCD calculation in
the hope that it will shed some light
on these quantities. 
 
The lattice formulation of the electromagnetic form factors has been given
in detail in the past~\cite{dwl90}. Here, we shall concentrate on the
DI contribution, where the strangeness current contributes.
In the Euclidean formulation, the Sachs EM form factors
can be obtained by the combination of two- and three-point functions
\begin{eqnarray}
\lefteqn{ G_{NN}^{\alpha\alpha}(t,\vec{p}) = \sum_{\vec{x}}e^{-i\vec{p}\cdot
\vec{x}}  \langle 0| \chi^\alpha(x) \bar{\chi}^\alpha(0) |0 \rangle }
\label{twopt} \\
\lefteqn{ G_{NV_{\mu}N}^{\alpha\beta}(t_f,\vec{p},t,\vec{q}) = \nonumber } \\
&& \sum_{\vec{x}_f,\vec{x}} e^{-i\vec{p}\cdot\vec{x}_f
  +i\vec{q}\cdot\vec{x}} \langle 0| \chi^\alpha(x_f) V_\mu(x)
  \bar{\chi}^\beta(0) |0 \rangle \label{threept},
\end{eqnarray}
where $\chi^\alpha$ is the nucleon interpolating field and $V_{\mu}(x)$
the vector current.
With large Euclidean time separation, i.e. $t_f - t >> a$ and $t >> a$,
where $a$ is the lattice spacing,
\begin{eqnarray}
\lefteqn{ \frac{\Gamma_i^{\beta\alpha}G_{NV_jN}^{\alpha\beta}(t_f,\vec{0},t,
\vec{q})} {G_{NN}^{\alpha\alpha}(t_f,\vec{0})}
\frac{G_{NN}^{\alpha\alpha}(t,\vec{0})}{G_{NN}^{\alpha\alpha}
(t,\vec{q})} } \nonumber \\
 && \longrightarrow  \frac{\varepsilon_{ijk}q_k}{E_q + m}
 G_M(q^2) \;, \label{mff} \\
\lefteqn{ 
  \frac {\Gamma_E^{\beta\alpha}G_{NV_4N}^{\alpha\beta}
      (t_f,\vec{0},t,\vec{q})} {G_{NN}^{\alpha\alpha}(t_f,\vec{0})}
  \frac {G_{NN}^{\alpha\alpha}(t,\vec{0})} {G_{NN}^{\alpha\alpha}
(t,\vec{q})} \longrightarrow   G_E(q^2) \label{eff} }
\end{eqnarray}
where $\Gamma_i= \sigma_i (1 + \gamma_4)/2$ and 
$\Gamma_E = (1 + \gamma_4)/2$.
 
We shall use the conserved current from the Wilson action which, being
point-split, yields slight variations
on the above forms and these are given in Ref.~\cite{dwl90}.
Our 50 quenched gauge configurations were generated on a $16^3 \times 24$
lattice at $\beta = 6.0$.
In the time direction,
fixed boundary conditions were imposed on the quarks to provide
larger time separations than available with periodic boundary
conditions. We also averaged over the directions of equivalent
lattice momenta in each configuration; this has the desirable effect of
reducing error bars.
Numerical details of this procedure are given in Refs. \cite{dwl90,ldd9495}.
The dimensionless nucleon masses $M_N a$ for
$\kappa = 0.154$, 0.152, and 0.148 are 0.738(16), 0.882(12), and
1.15(1) respectively. The corresponding dimensionless
pion masses $m_{\pi} a$ are
0.376(6), 0.486(5), and 0.679(4). Extrapolating the nucleon and
pion masses to the chiral limit we determine $\kappa_c = 0.1567(1)$
and $m_N a = 0.547(14)$.
Using the nucleon mass to set the scale to study
nucleon properties~\cite{ldd9495,dll9596},
the lattice spacing $a^{-1} = 1.72(4)$ GeV is determined. The
three $\kappa's$ then correspond to quark masses of about 120, 200,
and 360 MeV respectively.
 
The strangeness current $\bar{s}\gamma_{\mu}s$ contribution appears in the
DI only. In this case, we sum up the current insertion
$t$ from the nucleon source to the sink in Eqs.(\ref{mff})
and (\ref{eff}) to gain statistics~\cite{dll9596}.
The errors on the fit are
obtained by jackknifing the procedure.
To obtain the physical $G_M^s(q^2)$, we extrapolate the valence
quarks to the chiral limit while keeping the sea quark at the strange quark
mass (i.e. $\kappa_s$ = 0.154). It has been shown in the chiral
perturbation theory with a kaon loop that $G_M^s(0)$ is proportional to
$m_K$, the kaon mass~\cite{gss88}. Thus, we extrapolate
with the form $C + D \sqrt{\hat{m} + m_s}$ where $\hat{m}$ is the
average u and d quark mass and $m_s$ the strange quark mass to
reflect the $m_K$ dependence. This is the same form adopted for
extracting $\langle N|\bar{s}s|N\rangle$ in Ref.~\cite{dll9596},
which also involves a kaon loop in the chiral perturbation theory.

We obtain the extrapolated $G_M^s(q^2)$ at 4 nonzero $q^2$ values.
The errors are again obtained by jackknifing the extrapolation procedure
with the covariance matrix used to include the correlation among the
three valence $\kappa$'s. In view of the fact that the scalar current
exhibits a very soft form factor for the sea quark
(i.e. $g_{S,{\rm dis}} (q^2)$)
which has been fitted well with a monopole form~\cite{dll9596}, we shall
similarly use a monopole form to extrapolate $G_M^s(q^2)$ with nonzero
$q^2$ to $G_M^s(0)$. We find 
$G_M^s(0) = - 0.36 \pm 0.20 $.  Again, the correlation
among the 4 $q^2$ are taken into account and the error is from
jackknifing the fitting procedure. This is consistent with the
recent experimental value within errors (see Table 1).
We also find $G_{M,{\rm dis}}^{u/d}(0) =
 -0.65 \pm 0.30$, which is about 1.8 times the size of $G_M^s(0)$.
The sea contribution from the u, d, and s quarks $G_{M,{\rm dis}}^{u,d}(q^2)$
and $G_M^s (q^2)$ are added to the connected contributions
to give the full $G_M^p(q^2)$ and $G_M^n(q^2)$. 

   A similar analysis is done for the strange Sachs electric form factor
$G_E^s(q^2)$. We see that $G_E^s(0)$ is consistent with zero as it should
be and we find that the electric mean-square radius
$\langle r_s^2 \rangle_E = -6 dG_E^s(q^2)/dq^2|_{q^2 = 0} =
- 0.061 \pm 0.003\,{\rm fm}^2$. 
 
 \begin{table*}
\caption{Strangeness and proton-neutron m.\,m. and charge radii in
comparison with experiments.}
\vspace{0.2cm}
\begin{center}
\footnotesize
\begin{tabular}{llc}
 \multicolumn{1}{c}{} &\multicolumn{1}{c}{Lattice}
 & \multicolumn{1}{c} {Experiments} \\
 \hline
 $G_M^s(0)$ & $- 0.36 \pm 0.20 $ &
 $G^s_M(Q^2=0.1$GeV$^2)= 0.23\pm 0.37\pm 0.15\pm 0.19$~\cite{SAMPLE97} \\
 $G_{M,{\rm dis}}^u(0) $  & $- 0.65 \pm 0.30 $  &  \\
 $\mu_{{\rm dis}}$ & $- 0.097 \pm 0.037 \mu_N$  &   \\
 $\mu_p$ & $2.62 \pm 0.07\,\mu_N$  & $2.79\, \mu_N$ \\
 $\mu_n$  & $- 1.81 \pm 0.07\, \mu_N$ & $- 1.91\, \mu_N$  \\
 $\mu_n/\mu_p$ & $- 0.68 \pm 0.04$ & $- 0.685$ \\
 $\langle r_s^2 \rangle_E$ &  $- 0.061(3)$ --- $- 0.16(6)\, {\rm fm}^2$ &  \\
 $\langle r^2 \rangle_E^p$ &  $0.636\pm 0.046\,{\rm fm}^2$
 & 0.659 ${\rm fm}^2$~\cite{gal71}\\
 $\langle r^2 \rangle_E^n$ &  $- 0.123 \pm 0.019\,{\rm fm}^2$
 &  $- 0.127 \,{\rm fm}^2$~\cite{gal71}
 \\
 \hline
 \end{tabular}
 \end{center}
 \end{table*}

In summary, we have calculated the $s$ and $u$, $d$ contributions to
the electric and magnetic form factors of the nucleon. The individual m.\,m.
and electric form factors from the different flavors in the sea are not
small, however there are
large cancellations among themselves due to the electric charges of the
$u$, $d$, and $s$ quarks. We find
that a negative $G_M^s(0)$ leads to a total negative sea contribution to
the nucleon m.\,m. to make the
$\mu_n/\mu_p$ ratio consistent with the experiment.
We also find $G_E^s(q^2)$ positive which leads to a positive
total sea contribution to the neutron electric form factor $G_E^n(q^2)$.
Future calculations are needed to investigate the systematic errors
associated with the finite volume and lattice spacing as well as the
quenched approximation.

\noindent{\bf Acknowledgments:}
This work is partially supported by DOE
Grant DE-FG05-84ER40154 and by the Australian Research Council.

\end{document}